# Tailoring Interlayer Chiral Exchange by Azimuthal Symmetry Engineering

Yu-Hao Huang[1†], Jui-Hsu Han[1†], Wei-Bang Liao[1], Chen-Yu Hu[1], Yan-Ting Liu[1], and Chi-Feng Pai[1,2,3*]

[1]*Department of Materials Science and Engineering, National Taiwan University, Taipei 10617, Taiwan*

[2]*Center of Atomic Initiative for New Materials, National Taiwan University, Taipei 10617, Taiwan*

[3]*Center for Quantum Science and Engineering, National Taiwan University, Taipei 10617, Taiwan*

## Abstract

Recent theoretical and experimental studies of the interlayer Dzyaloshinskii-Moriya interaction (DMI) has sparked great interest in its implementation into practical magnetic random-access memory (MRAM) devices, due to its capability to mediate long-range chiral spin textures. So far, experimental reports focused on the observation of interlayer DMI, leaving the development of strategies to control interlayer DMI's magnitude unaddressed. Here, we introduce an azimuthal symmetry engineering protocol capable of additive/subtractive tuning of interlayer DMI through the control of wedge deposition of separate layers, and demonstrate its capability to mediate field-free spin-orbit torque (SOT) magnetization switching in both orthogonally magnetized and synthetic antiferromagnetically coupled systems. Furthermore, we showcase the spatial inhomogeneity brought about by wedge deposition can be suppressed by specific azimuthal engineering design, ideal for practical implementation. Our findings provide guidelines for effective manipulations of interlayer DMI strength, beneficial for future design of SOT-MRAM or other spintronic devices utilizing interlayer DMI.

[†] Y.-H. H. and J.-H. H. contributed equally to this work
[*] Email: cfpai@ntu.edu.tw

The spin-orbit torque (SOT) originating from the spin current generated from either the spin-Hall effect or the Rashba effect, has the capability to electrically manipulate magnetic moments. SOT based magnetic random-access memory (SOT-MRAM), compared to conventional spin-transfer torque MRAM, enjoys key advantages such as higher operation speed[1, 2], improved endurance[3, 4], and potentially higher charge-to-spin conversion efficiency[5-7]. Currently, SOT-MRAM based on magnetization with perpendicular magnetic anisotropy (PMA) has been considered a prime candidate for the next generation non-volatile random-access memory, due to its combination of high storage density and good thermal stability[4, 8-10]. However, due to symmetry constraint, an in-plane polarized spin current cannot switch the PMA magnetization deterministically without an external in-plane field[11, 12]. Numerous solutions to this conundrum have been proposed, including the use of lateral symmetry breaking[13-16], exchange coupling[17-20], unconventional spins[7, 21, 22] or geometry engineering[23]. These methods often generate new issues such as incompatibility with established magnetic tunnel junction (MTJ) fabrication processes, low magnetization switching ratio[19, 24], or low spin-conversion efficiency[22].

Recent calculations[25, 26] and experimental verifications[27-32] have broadened the horizons of antisymmetric Dzyaloshinskii-Moriya interaction (DMI)[33-36] from the conventional interfacial case[37-39] to a long range interlayer scenario. Identical to the interfacial DMI, interlayer DMI requires high spin-orbit coupling (SOC), with additional symmetry-breaking element introduced in the *xy* plane. Under interlayer DMI, chiral spin configuration exists between two discrete magnetic layers separated by a spacer rather than at the ferromagnet (FM)/Normal metal (NM) interface, enabling novel magnetic configurations manipulation. Previous works showcased the fundamental importance of wedge (oblique) deposition as the in-plane symmetry breaking element to generate interlayer DMI [26, 28, 40] and assist field-free current-driven magnetization switching[41, 42]. Modelling of the Fert-Lévy three-site model also revealed the oscillatory nature of interlayer DMI [26]. However, little attention had been put into the strength control of interlayer DMI, and how the interplay of wedge deposition's detailed conditions serves as an important tuning parameter of interlayer DMI. Additionally, unfavorable device-to-device characteristic deviations that may accompany the wedge deposition process remain unaddressed.

Inspired by successful utilization of the oblique angle deposition (OAD) technique in various scientific domains[43-45], this work experimentally reveals how using OAD to create different wedge deposited layers leads to opposite contributions to the overall strength of interlayer DMI, in both orthogonally magnetized (type-T) and perpendicularly-magnetized synthetic antiferromagnet (SAF) systems. This is qualitatively explained by a toy model based on Fert's three-site picture. Accordingly, we introduce an antiparallel wedge-deposition technique dubbed "counter-wedge", and explore how enhancing the interlayer DMI effective field ($H_{DMI}$) is enabled by azimuthal symmetry modulation. Additionally, a significant $H_{DMI}$ is observed even in the "self-counter-wedge" condition (*i.e.* counter-wedge within a single layer) and robust field-free current-induced switching can be achieved in both SAF and type-T systems. Finally, suppression of device-to-device variation is proven by comparing the switching characteristics between samples grown by counter-wedge and self-counter-wedge. Our discoveries of $H_{DMI}$ enhancement through azimuthal symmetry engineering and the corresponding field-free magnetization switching with improved device-to-device homogeneity may advance the practical potential of SOT-MRAM devices.

Two series of magnetic multilayers, S1: Ta(0.5)/Pt(2)/Co(0.75)/Pt(2)/CoFeB(1.35)/Ta(2) (number in the parentheses in nm) and S2: Ta(0.5)/[Pt(1)/Co(0.4)]$_3$/Ru(0.5)/[Co(0.4)/Pt(1)]$_2$/Ta(2) are deposited on thermally-oxidized Si/SiO$_2$ substrate using high-vacuum magnetron sputtering (see Methods). Pt/Co based structures are chosen for its well-known PMA characteristics and high SOC at the interface, satisfying the requirements for DMI generation (though, a recent publication successfully incorporated spacer with strong Rashba coupling[46]). S1 possesses type-T magnetic configuration with Co exhibiting PMA due to strong SOC at the Pt/Co and Co/Pt interfaces while the relatively thick CoFeB is in-plane magnetized. An optimized Ru thickness of 0.5 nm is chosen[47, 48] to induce Ruderman-Kittel-Kasuya-Yosida (RKKY) interaction that couples [Pt(1)/Co(0.4)]$_3$ and [Co(0.4)/Pt(1)]$_2$ layers antiferromagnetically, thereby forming a perpendicular synthetic antiferromagnet (SAF) in S2. For both series, individual Pt layers are selectively wedge deposited to generate in-plane symmetry breaking. The films are patterned into micron-sized Hall bar devices for further characterizations. Afterwards, several other stack designs are devised based on the insights gained from S1 and S2.

We start by investigating different wedge layer's contribution to the overall interlayer DMI strength in S1. To specify, the wedge growth is achieved by OAD with the Pt atom flow having an incident angle of 20° relative to the substrate normal with an azimuthal angle of 0° (parallel to the $x$ axis). A total of four wedge scenarios are created, first being Pt buffer layer (designated $Pt_1$) and spacer layer (designated $Pt_2$) both grown with OAD. The second and third scenario involves only $Pt_1$ or $Pt_2$ grown with OAD, respectively. In the last case, $Pt_1$ is again deposited with standard OAD, but during the sputtering of $Pt_2$, the holder is rotated by 180° with regard to $Pt_1$'s wedge direction (azimuthal angle being 180°), creating the counter-wedge scenario. Figure 1a provides graphical representation of the counter-wedged structure and inset in Figure 1d sketches critical layers for all scenarios. In a system where magnetizations $\mathbf{M}_1$ and $\mathbf{M}_2$ are coupled by interlayer DMI, their energy density is expressed as $\widehat{E}_{DMI} = -\mathbf{D} \cdot (\mathbf{M}_1 \times \mathbf{M}_2)$, where $\mathbf{M}_1$ and $\mathbf{M}_2$ denotes magnetizations' unit vectors. Under a symmetry breaking factor controlled by the Pt layer(s) atom flow direction, the vector $\mathbf{D}$ characterizes the strength and direction of interlayer DMI, and should lie within the $xy$ plane[30, 49], perpendicular to said symmetry breaking to satisfy the third Moriya rule[34]. In the case of S1 samples where the two magnetizations are orthogonal, by taking $\mathbf{M}_1$ and $\mathbf{M}_2$ to be $\mathbf{M}_{Co}$ and $\mathbf{M}_{CoFeB}$ respectively, $\widehat{E}_{DMI}$ gives rise to an effective DMI field $\mathbf{H}_{DMI} = -\mathbf{D} \times \mathbf{M}_{CoFeB}$ acting on $\mathbf{M}_{Co}$ with $\mathbf{M}_{CoFeB}$ reciprocally experiencing an effective field of $\mathbf{D} \times \mathbf{M}_{Co}$. To $\mathbf{M}_{Co}$ possessing PMA, $\mathbf{H}_{DMI}$ points purely in the $z$ direction owing to the in-plane oriented $\mathbf{M}_{CoFeB}$, and the relative orientation between $\mathbf{D}$ and $\mathbf{M}_{CoFeB}$ dictates the polarity and magnitude of $\mathbf{H}_{DMI}$.

Accordingly, angle dependent loop-shift measurement is devised to characterize interlayer DMI strength[30, 41, 42]. An in-plane field $\mathbf{H}_{in}$ ($|H_{in}|$ = 200 Oe) is applied to prefix $\mathbf{M}_{IMA}$, while the out-of-plane field $H_z$ is swept to capture the hysteresis behavior of the PMA Co layer through anomalous

Hall resistance ($R_H$) measurement on the Hall bar devices (schematics in Figure 1a lower right panel). A typical $R_H$-$H_z$ loop of the counter-wedged sample demonstrating significant hysteresis loop offset depending on in-plane field's azimuthal direction $\varphi_H$ is shown in Figure 1b, quantified by the shift in the $R_H$-$H_z$ loop centers, namely $H_{\text{offset}}$. By gradually rotating $\varphi_H$ from 0° to 360°, $H_{\text{offset}}$ as a function of $\varphi_H$ is summarized in Figure 1c. The observed clear antisymmetric behavior of $H_{\text{offset}}$ confirms the strong interlayer DMI within S1 samples, from which their $\mathbf{H}_{\text{DMI}}$ magnitude, $H_{\text{DMI}}$ and angular dependences can be extracted by sine fit $H_{\text{offset}} = H_{\text{DMI}} \sin(\varphi_H - \varphi_D)$ ($\varphi_D$ representing $\mathbf{D}$'s azimuthal angle). In the case of wedged spacer Pt$_2$, sine fit indicates $\varphi_D \approx 286°$, in good agreement with the constraint that $\mathbf{D}$ must be perpendicular to the symmetry breaking vector and a $H_{\text{DMI}}$ of 25.5 Oe is obtained. On the other hand, when the buffer Pt$_1$ is wedged, while still following the third Moriya rule, $\mathbf{D}$ is rotated by ~ 180° compared to its Pt$_2$-wedged counterpart, at $\varphi_D \approx 99°$, simultaneously possessing a high $H_{\text{DMI}} = 87.1$ Oe.

Judging from these two results, the same OAD conditions employed on different wedged layers appear to generate opposite symmetry breaking polarities, consequently revealing the possibility of azimuthal-symmetry engineering. In the full wedge sample (both Pt$_1$ and Pt$_2$ are prepared by OAD with azimuthal angle of 0°), $H_{\text{DMI}}$ is weakened to 68.6 Oe while maintaining identical $\varphi_D$ as the Pt$_1$ wedge scenario. In contrast, the counter-wedge sample (Pt$_1$-0° with Pt$_2$-180°) shows an enhancement of $H_{\text{DMI}}$ to 113.8 Oe. These results are compiled in Figure 1d. Reminiscent to the fact that the full wedge structure can be viewed as combining Pt$_1$ wedge and Pt$_2$ wedge, its $H_{\text{DMI}}$ also matches with the mutual cancellation due to the their antiparallel $\varphi_D$. A conversed scenario holds for the counter-wedge case, in which by turning over Pt$_2$ wedge's corresponding $\varphi_D$ through reversed OAD, subtraction of $H_{\text{DMI}}$ becomes addition.

From a microstructure's point of view, OAD is often considered responsible for creating slanted columnar structures due to the shadow effect[45, 50], where previous experimental studies have correlated microstructures or textures induced field-free SOT switching to OAD via scanning electron microscopy[51] and X-ray diffraction pole figure measurements[52]. Such effect could potentially create horizontal shifts of the atoms in the Co and the CoFeB layers due to template effect from the wedged Pt nanostructures. Subsequently, a toy model is established by incorporating these horizontal shifts into the Fert-Lévy three-site model, a strategy also employed by previous works[25, 26], with our model focusing on the effects both the Pt buffer and spacer layers exert onto the two magnetic layers (See Supporting Information I for detailed descriptions). Calculations from this toy model predicts a superposition relationship between the overall magnitude of **D** and contributions from individual symmetry-breaking elements, supporting our observed additive/subtractive engineering results of $H_{DMI}$ (template scenario explains how wedged Ta buffer layer induces sizable interlayer DMI in a previous report[42] while said Ta layer is distant from the spacer layer). Intriguingly, a similar model predicts the oscillatory nature of interlayer DMI[26] with experimental evidence[26, 53], solidifying our proposed model's capability at capturing vital characteristics of interlayer DMI. We also comment that possible thickness gradient due to OAD is unlikely the source of interlayer DMI since thickness difference within single 50-μm long device is a miniscule $1.6 \times 10^{-3}$ nm. This gradient becomes non-negligible when the area-of-interest is enlarged to wafer-scale (thus we propose the self-counter wedge solution as introduced later), but does not significantly influence the out-of plane anisotropy field in our case, as discussed in Supporting Information II.

We further investigate how manipulating the wedge conditions can influence interlayer DMI in S2 samples (SAF structures). Figure 2a shows the SAF structure, where the enlarged illustration giving an example of the wedge conditions of various Pt layers (in this case, counter-wedge). The inset in Figure 2b shows the full $R_H$-$H_z$ loop with a square hysteresis near the origin. Clearly, the bottom [Pt/Co]$_3$ (abbreviated as **M**$_B$) and the upper [Co/Pt]$_2$ (abbreviated as **M**$_T$) is antiferromagnetically coupled by the Ru(0.5) spacer, and can be forced to a parallel state under high $H_z$. In S2, due to the $z$-orientated nature of **M**$_B$ and **M**$_T$, $H_{in}$ now creates in-plane tilt of the PMA magnetizations to generate nonzero $H_{offset}$ in the form of $\mathbf{H}_{DMI(T)} = -\mathbf{D} \times \mathbf{M}_B$ ($\mathbf{H}_{DMI(B)} = \mathbf{D} \times \mathbf{M}_T$) with regard to **M**$_T$ (**M**$_B$)[30, 40]. Figure 2b shows that $H_{in}$ breaks the symmetry of magnetic hysteresis, with antisymmetric $H_{offset}$ clearly observed, serving as a good figure of merit for comparing strength and direction of interlayer DMI.

The three Pt layers located in $\mathbf{M_B}$, designated as $Pt_1$, $Pt_2$ and $Pt_3$ (see Figure 2a), are further scrutinized to see their individual contributions to the strength of interlayer DMI. Deposited using standard OAD, angle dependent loop-shift results for $Pt_1$, $Pt_2$ and $Pt_3$-wedged samples are shown in Figure 2c. Their angular dependences clearly share the identical symmetric/antisymmetric axis with the results reported in S1, where sinusoidal fits can be used to extract $H_{DMI}$ and $\varphi_D$. In the case where $Pt_1$ is grown with OAD, $\varphi_D \approx 271°$ with $H_{DMI} = 77.9$ Oe. On the contrary, wedged $Pt_2$ and $Pt_3$ results in an antiparallel $\varphi_D$ of 91°, with $H_{DMI}$ of 71.0 Oe and 38.9 Oe, respectively. Under the picture of Fert's three-site model with shifted atomic sites (Supporting Information I), it's implied that wedged $Pt_1$ serves to displace the magnetic atoms in $\mathbf{M_B}$, and wedged $Pt_2$ and $Pt_3$ affect the displacements in $\mathbf{M_T}$, by taking analogy with the opposite contribution from the Pt buffer and spacer in S1 (type-T). Importantly, by combining OAD in $Pt_2$ and $Pt_3$ with a reversed OAD in $Pt_1$ (denoted counter-wedge), an increased $H_{DMI}$ of 104.0 Oe is also observed. While this value is not a precise quantitative summation of individual $H_{DMI}$ contributions, the profound qualitative enhancement when compared to the reduction under the destructive full-wedge situation again shows the viability of azimuthal engineering of interlayer DMI, as summarized in Figure 2d. Additionally, observations on S1 and S2 samples unequivocally point out that the earlier a wedged layer is grown onto the substrate, the higher an influence it has on the interlayer DMI strength, consistent with the picture of template effect and columnar structure formation.

Aside from changing wedge combinations for interlayer DMI strength engineering, another previously unexplored pathway is to change the deposition angle when executing OAD, thereby changing the magnitude of symmetry breaking as well. For both type-T and SAF systems, raising the OAD's incident angle from 20° (used in the main text) to 70° results in a up to four-fold increase in $H_{DMI}$, as elaborated in Supporting Information III (damped oscillatory $H_{DMI}$-spacer thickness behavior also captured under 50° OAD incident angle). This method alongside the proposed wedge manipulation technique have the potential to be combined to better fine-tune the strength of interlayer chiral exchange as the fabrication processes see fit.

Having revealed the controllability of interlayer DMI strength via azimuthal engineering, we demonstrate field-free current-induced SOT switching using the two counter-wedge deposited samples from S1 and S2. Hall bar devices with various orientations are utilized. This geometry creates a pivoting $\varphi_D$ with regard to the *x* axis of individual devices (Figure 1 a, b) to identify switching's angle dependence. Switching experiments are performed at a magnet-free probe station. As shown in Figure 3a and b, robust and deterministic field-free current-driven switching in both type-T and SAF can be achieved. In both cases, the switching polarity and the switching percentage are strongly correlated to the interplay between ***D*** and charge (spin) current. Under a sizable charge current $I_{pulse}$ along the current channel (*x* axis), magnetization dynamics are activated by a conventional *y*-polarized spin (σ) and eventually aligns **M**$_T$ and **M**$_B$ collinearly toward *y* due to the current-induced SOT[54] (See Supporting Information IV). In other words, σ induces an in-plane component in both magnetizations. **M**$_T$ and **M**$_B$ therefore experience **H**$_{DMI}$ acting as a *z*-directional assist field proportional to σ × **D** and **D** × σ (recall **H**$_{DMI(T)}$ = −**D** × **M**$_B$ and **H**$_{DMI(B)}$ = **D** × **M**$_T$), respectively. As illustrated in Figure 3c, $\varphi_D$ = 0° (180°) minimizes (maximizes) σ × **D**, resulting in **M**$_T$ deterministically stabilizing to ± *z* directions, while **M**$_B$ relaxes to an opposite direction with regard to **M**$_T$ (due to the antisymmetric **H**$_{DMI}$) thus completing a deterministic switching cycle. Such simultaneous switching of the two magnetizations has been experimentally examined in a previous work on type-T structure[41], where the current-induced switching of the in-plane magnetized CoFeB exerts $H_{DMI}$ that directly switches the PMA magnetization. On the contrary, $\varphi_D$ = 90° and 270° results in a null σ × **D**, and failed switching are experimentally confirmed (Figure 3a, b) due to the diminishingly small *z*-directional assist fields. The switching percentages as functions of $\varphi_D$ are summarized in Figure 3d. The maximum switching percentages (92% and 94% for type-T and SAF cases, respectively) occurred at $\varphi_D$ = 0° and 180°, consistent with the fact that $\varphi_D \| I_{pulse}$ maximizes the *z*-oriented assist fields. This comparatively high switching percentage is an improvement over previous reports, possibly due to the enhanced $H_{DMI}$, which overcomes the depinning fields or the Oersted fringing fields[30, 49].

Azimuthal engineering presents opportunities for achieving high $H_{DMI}$ and field-free switching. In addition, counter-wedge technique is an improvement over conventional wedge deposition since nominally, the antiparallel thickness gradient in discrete Pt layers cancel out with each other. However, counter-wedge suffers from a drawback where thickness mitigation is only achieved after the growth of multiple layers, potentially generating slanted layer stacks (as illustrated in Figures 1a and 2a). We hereby propose another improvement dubbed "self-counter wedge", in which individual Pt layers are sequentially grown by an OAD and a reversed OAD. Via this approach, the thickness gradient is tentatively smoothened out within a single Pt layer. In type-T, self-counter wedge employed on $Pt_1$ results in $H_{DMI}$ reduction to a smaller yet robust $H_{DMI}$ of 51.8 Oe (Supporting Information V). $\varphi_D \approx 270°$ signifies the dominant role of the lower half of the Pt layer, overcompensating for the upper half of Pt layer neighboring the Co layer that's deposited with comparatively reversed atom flow direction (which contribute to $\varphi_D \approx 90°$), in concert with previous observations that the template effect profoundly governs the strength and the direction of **D**. Figure 4a showcases the SAF structure with $Pt_1$, $Pt_2$ and $Pt_3$ grown by the self-counter wedge protocol. Highlighted in the enlarged image, the OAD and the reversed OAD sequences employed are opposite for $Pt_2$ and $Pt_3$ when compared to $Pt_1$ to take advantage of the overall additive contributions. The self-counter wedge process provides $H_{DMI} \approx 52.5$ Oe in the SAF devices (Supporting Information V). The individual contributions from $Pt_1$, $Pt_2$ and $Pt_3$ layers detailed in Supporting information VI. Their identical resulting **D** directions and good $H_{DMI}$ agreement with the overall structure indicate the controllability and predictability of azimuthal symmetry engineering of interlayer DMI in SAF structures. In addition, wedged Pt layers in the top FM layer result in meager $H_{DMI}$ at 5.7 Oe, in line with the weaker template efficacy consistently observed in later-deposited layers. To compare the uniformity of self-counter and counter wedge grown structures, SAF devices are deposited on a 3-inch (7.6-cm) wafer as shown in Figure 4b where five zones with 1-cm spacings are chosen to test their spatial-dependent device characteristics. Resistivity measurement results shown in Figure 4b demonstrate electrical property's spatial homogeneity, where the resistivity's spatial variation is suppressed to under ± 5% in the case of self-counter wedge (± 20% for the counter case), possibly due to decreased interface roughnesses[55]. Magnetic switching characteristics are also scrutinized. As shown in Figure 4c, robust field-free switching is achievable for self-counter wedged devices in all five zones. Normalized $\Delta R_H$ and critical switching current $I_c$ of the tested devices are compiled in Figure 4d, e, respectively. For the counter-wedge case, the switching percentage varies wildly, possibly due to the slanted multilayers phenomenon exacerbated by the devices' spatial deviation from the nominal sputtering position. In the self-counter case, the switching percentage is more consistent across these five zones. A similar scenario is identified when characterizing $I_c$, where the self-counter wedged samples maintain a fairly stable $I_c$, while for the counter-wedge case the $I_c$ variation exceeds 5 mA among different regions. These results demonstrate that the self-counter wedge process is a feasible approach to mitigate the spatial inhomogeneity issue, showing its potential for practical utilization.

In conclusion, we reveal the overall strength and direction of interlayer DMI are governed by the individual contributions from separate wedged Pt layers, in both orthogonally magnetized (type-T) and synthetic antiferromagnetic (SAF) structures. These contributions, while strictly following the Moriya symmetry rules, possess opposite polarities that can be subsequently harnessed by our proposed method for tuning/maximizing interlayer DMI strength with adequately designed wedge deposition configurations. These results are qualitatively supported by a toy model based on the Fert-Lévy model. Subsequently, successful field-free current-induced SOT switching is achieved in both type-T and SAF systems. Finally, a "self-counter wedge" method is proposed to achieve improved homogeneity in field-free switching characteristics, albeit with the trade-off of reduced interlayer DMI effective fields. Overall, our findings provide further insights to the microscopic origin of interlayer DMI, and form strategies to better manipulate 3D spin textures, therefore paving the way for interlayer DMI's potential utilization in next generation spintronic devices.

**Materials and methods**

The magnetic multilayer stacks are prepared by DC magnetron sputtering under a base pressure of ~ $1\times10^{-8}$ Torr, and Ar working pressure of $3\times10^{-3}$ Torr. Throughout this work, CoFeB is the abbreviation for $Co_{40}Fe_{40}B_{20}$ (subscript denotes the atomic percentage). The Pt layers are selectively sputtered at an oblique incidence angle of 20° except when otherwise noticed (varied incident angle investigations conducted in the Supporting Information III), with all other layers grown with sample rotation enabled to ensure homogeneity. Hall bar devices with 5 μm width and 50 μm length are then prepared with photolithography and lift-off process. A well calibrated vector magnet system (Model 5204, GMW Associates) is used to simultaneously provide in-plane and out-of-plane fields, while a Keithley 2400 sourcemeter and a Keithley 2000 multimeter provides DC current and measures Hall signal, respectively. Field-free current-induced magnetization switching is performed at a magnet-free station, with the Hall signals captured by an identical electrical setup. All measurements are performed at room temperature.

**Supporting information**

A toy model to demonstrate the relative contribution from separate wedged Pt layers and their combined effect; deposited film thickness gradient and anisotropy field across a distance; correlation of OAD's incident angle with resultant $H_{DMI}$ in structures; measurement of SOT efficiency for type-T and SAF samples, thermal stability and switching probability of the SAF sample; type-T self-counter wedged $H_{DMI}$ and field-free switching demonstrations; details of individual layer's contribution in the self-counter wedge scenario.

**Acknowledgments**

This work is supported by the National Science and Technology Council (NSTC) under grant No. NSTC 112-2636-M-002-006 and grant No. 112-2112-M-002-031. We would also like to acknowledge the support from the Center of Atomic Initiative for New Materials (AI-Mat) and the Advanced Research Center of Green Materials Science and Technology, National Taiwan University from the Featured Areas Research Center Program within the framework of the Higher Education Sprout Project by the Ministry of Education (MOE) in Taiwan under grant No. NTU-112L9008.

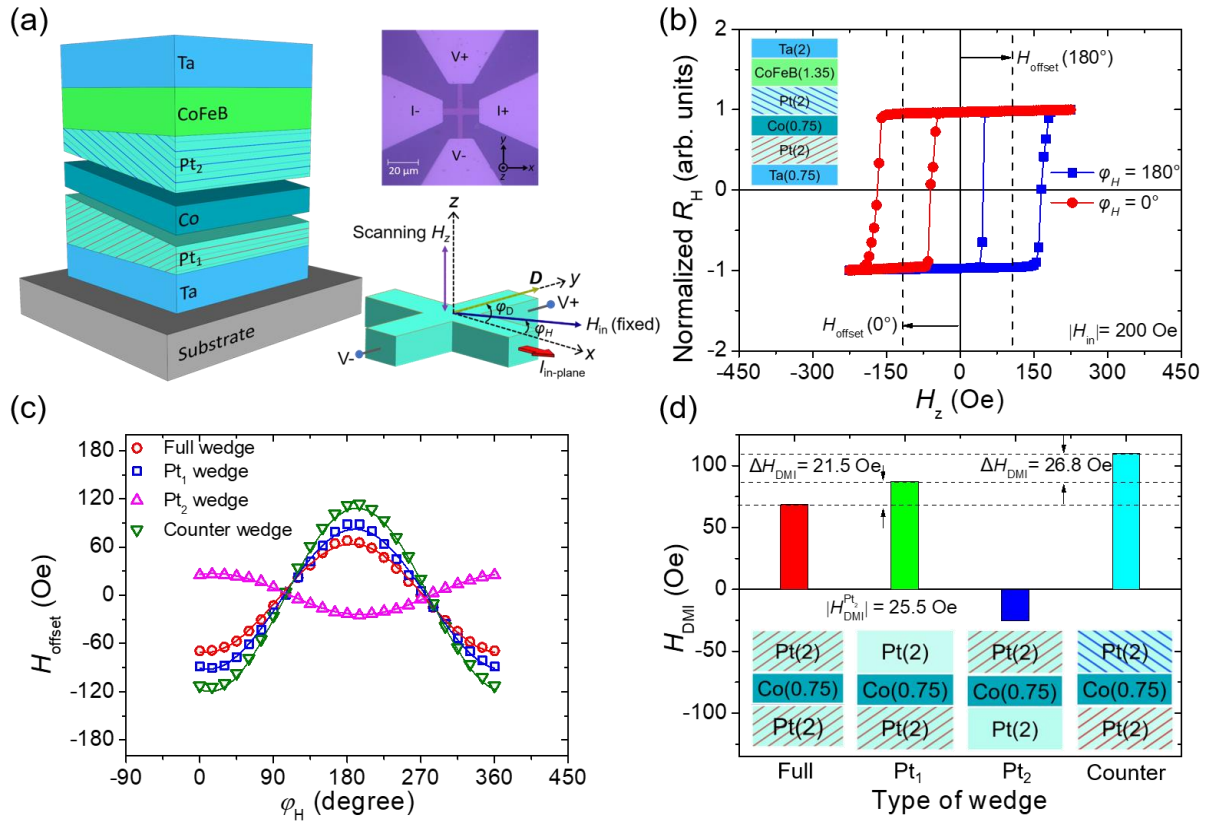

Figure 1. Characterization of interlayer DMI induced effective fields $H_{DMI}$ in type-T samples (Series S1) with different wedge conditions. (a) Left: illustration of a representative counter-wedge grown sample. $Pt_1$ and $Pt_2$ denoting the buffer and the spacer Pt layers, respectively. Counter-wedge graphically highlighted by red/blue stripes (guide to the eye) to emphasize its origin from the anti-parallel atom flow directions. The atom flow points toward the stripes' plane normal. Upper right: Optical microscope image of a representative Hall bar device. Lower right: angle dependent loop-shift measurement setup. (b) Representative $R_H$ vs. $H_z$ loops of a counter-wedge sample measured with $\varphi_H = 0°$ and $\varphi_H = 180°$. (c) The offset field $H_{offset}$ as a function of $\varphi_H$ for different types of wedge growth conditions. Solid lines are sine fits to the data, from which $H_{DMI}$ are extracted. (d) $H_{DMI}$ for different wedge types. Inset shows their respective 2D critical layers representations (Full, 1st Pt, 2nd Pt and counter-wedge, respectively). The magnitude of $H_{DMI}$ is designated as negative for 2nd Pt wedge as a basis of comparison.

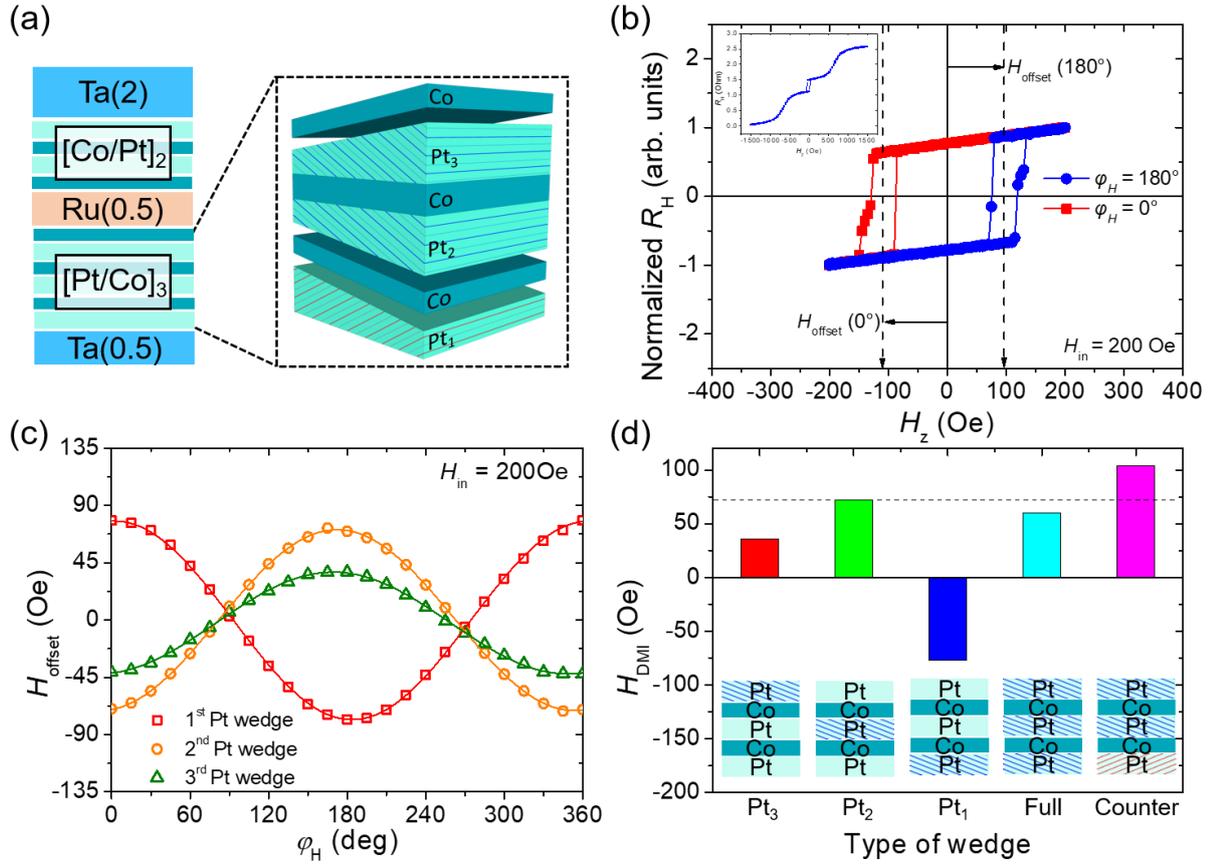

Figure 2. Characterization of $H_{DMI}$ in SAF structures (Series S2) with various wedge deposition conditions. (a) Depiction of SAF stack's structure, the enlarged image details the bottom $[Pt/Co]_3$, showing the "counter-wedge" configuration. (b) Asymmetric hysteresis of the central loop with $|H_{in}|$ = 200 Oe under $\varphi_H$ = 0° and 180°. (c) Angle scan for $H_{offset}$ of $Pt_1$, $Pt_2$ and $Pt_3$ wedged samples. Solid lines are sine fits to the data. (d) $H_{DMI}$ compilation of samples with different wedge combinations, similar to the case in type-T structures, the magnitude of $H_{DMI}$ in wedged $Pt_1$ is designated negative. Inset details different wedge conditions.

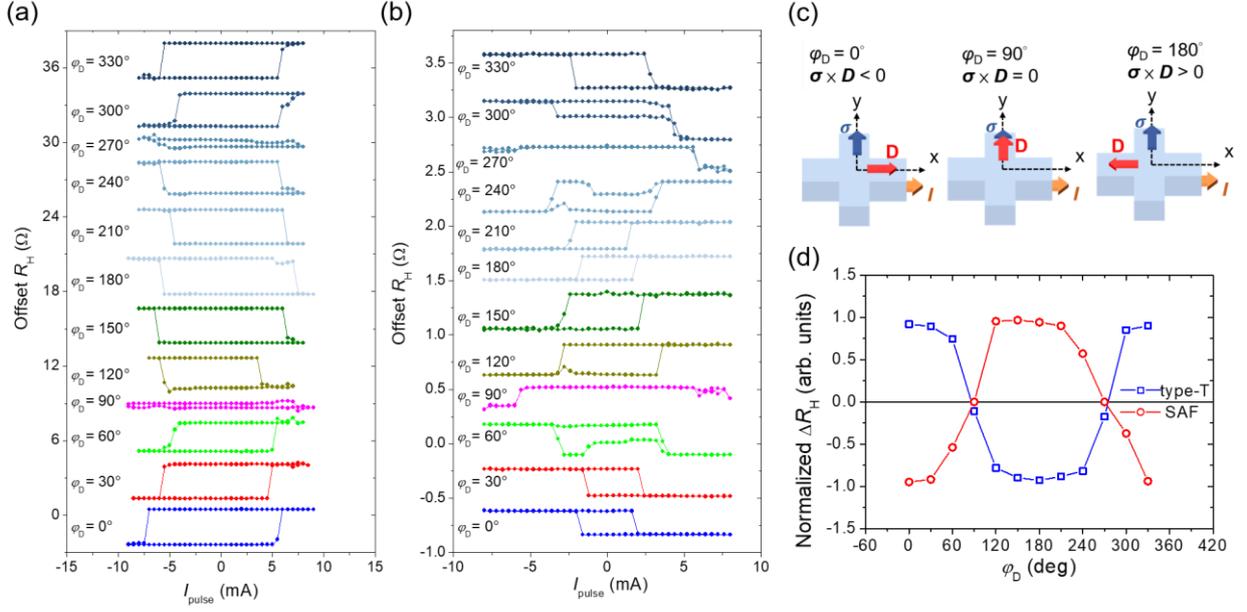

Figure 3. Demonstration of field-free current-induced SOT switching assisted by counter-wedge induced interlayer DMI. Field-free SOT switching behaviors of (a) type-T counter-wedged and (b) SAF counter-wedged devices with various $\varphi_D$. Note the change in polarity with regard to the $\varphi_D$ evolution, and failed switching occurring at $\varphi_D = 90°, 270°$. (c) Depiction of the field-free switching mechanism. (d) Switching percentage (demonstrated by normalizing to the $R_H$ obtained from field-scan) for both types of samples as a function of $\varphi_D$.

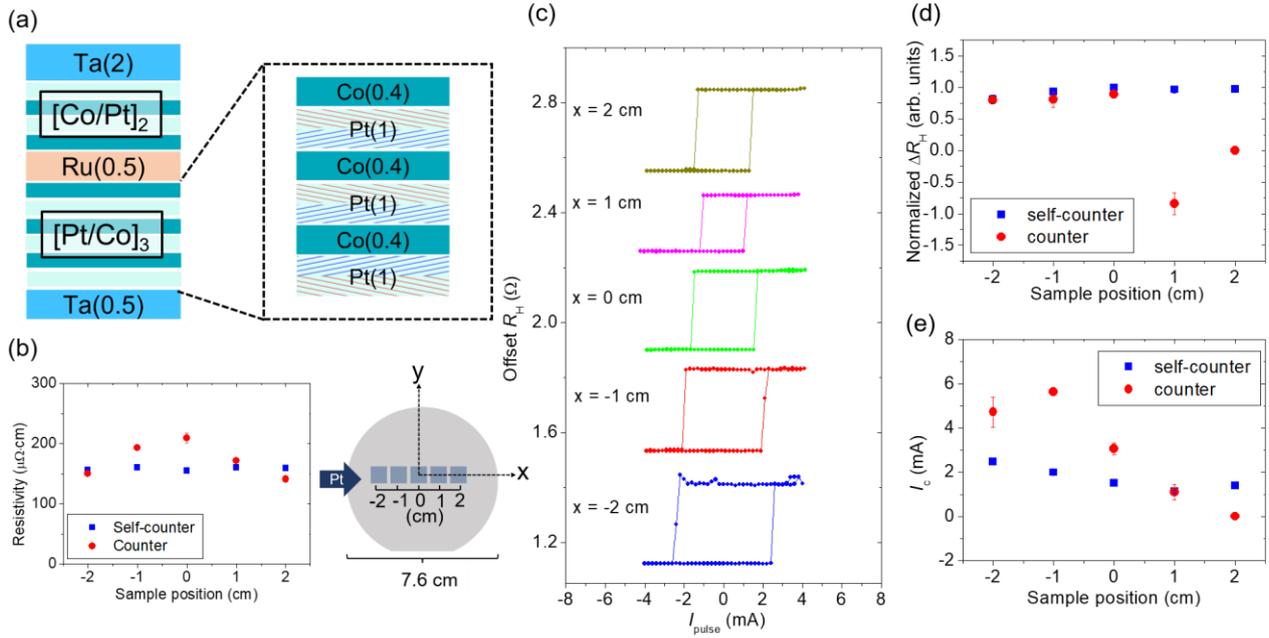

Figure 4. Improved field-free switching homogeneity by employing self-counter wedge. (a) Structure of the self-counter wedged SAF stack with the wedged induced non-uniformity being mitigated in individual layers. Enlarged section highlights the opposite order of OAD and reversed OAD for the three Pt layers involved. (b) Devices deposited on a 3-inch (7.6-cm) wafer and five zones are chosen (with spacings of 1 cm) along the Pt atom flow direction (*x* axis) to test their field-free switching characteristics. Resistivity as a function of sample position on the substrate is also shown. (c) Field-free switching loops of $\varphi_D = 180°$ devices from the five zones. (d), (e) Compares the switching percentage (normalized) $\Delta R_H$ and critical switching current $I_c$, respectively, of devices with $\varphi_D = 180°$ from different zones between the two samples.

Supporting Information for

# Tailoring Interlayer Chiral Exchange by Azimuthal Symmetry Engineering

Yu-Hao Huang[1†], Jui-Hsu Han[1†], Wei-Bang Liao[1], Chen-Yu Hu[1], Yan-Ting Liu[1], and Chi-Feng Pai[1,2,3*]

[1]*Department of Materials Science and Engineering, National Taiwan University, Taipei 10617, Taiwan*

[2]*Center of Atomic Initiative for New Materials, National Taiwan University, Taipei 10617, Taiwan*

[3]*Center for Quantum Science and Engineering, National Taiwan University, Taipei 10617, Taiwan*

**Contents**

**Note I.** A toy model of interlayer DMI based on the Fert-Lévy three-site scenario

**Note II.** Thickness gradient and out-of-plane anisotropy field across a distance

**Note III.** OAD incident angle dependence of $H_{\mathrm{DMI}}$

**Note IV.** Damping-like SOT efficiency characterization, thermal stability and switching probability characterization for the SAF sample

**Note V.** $H_{\mathrm{DMI}}$ characterization and current-induced field-free switching in a type-T self-counter-wedged sample

**Note VI.** Individual layer's contribution in the self-counter wedge scenario, and weak $H_{\mathrm{DMI}}$ from wedged Pt layers in the top FM layer

---

* Email: cfpai@ntu.edu.tw

**Note I. A toy model of interlayer DMI based on the Fert-Lévy three-site scenario**

A toy model consisting of two connected Fert-Lévy three-site triangles, as illustrated in Fig. S1(a), is examined. Following the procedures described in[25, 26], DMI vectors from the two triangles are written as

$$\boldsymbol{D}_1(\boldsymbol{R}_{li}, \boldsymbol{R}_{lj}, \boldsymbol{R}_{ij}) = -V_1 \frac{(\boldsymbol{R}_{li} \cdot \boldsymbol{R}_{lj})(\boldsymbol{R}_{li} \times \boldsymbol{R}_{lj})}{|\boldsymbol{R}_{li}|^3 |\boldsymbol{R}_{lj}|^3 |\boldsymbol{R}_{ij}|} \qquad (S1)$$

and

$$\boldsymbol{D}_2(\boldsymbol{R}_{l'i}, \boldsymbol{R}_{l'j}, \boldsymbol{R}_{ij}) = -V_1 \frac{(\boldsymbol{R}_{l'i} \cdot \boldsymbol{R}_{l'j})(\boldsymbol{R}_{l'i} \times \boldsymbol{R}_{l'j})}{|\boldsymbol{R}_{l'i}|^3 |\boldsymbol{R}_{l'j}|^3 |\boldsymbol{R}_{ij}|} \qquad (S2)$$

where various $\boldsymbol{R}$ vectors represent interatomic vectors for the left and right Fert-triangles, respectively, and $V_1$ is a material-dependent quantity. In the absence of symmetry breaking, the total magnitude of $\boldsymbol{D}$ (sum of $\boldsymbol{D}_1$ & $\boldsymbol{D}_2$) is naturally cancelled out (dashed lines connected triangles in Fig. S1(a)). However, when atomic displacements (labeled $\delta_1$ and $\delta_2$) occur in the lower Co or the upper CoFeB layers (due to wedge deposition along $x$ axis), the magnitude of $\boldsymbol{D} = \boldsymbol{D}_1 + \boldsymbol{D}_2$ becomes nonvanishing.

$\boldsymbol{D}$'s total magnitude is calculated by the summation of eq. S1 and eq. S2. To demonstrate a more semi-quantitative agreement with the experimental results, displacements (represented by shifts $\delta_1$ and $\delta_2$) are tentatively set as 8% and 2.4% of the interatomic distance, respectively. Consequently, individually taking $\delta_1$ or $\delta_2$ into consideration results in nonvanishing $\boldsymbol{D}$ with the same polarity (toward $+y$), and their corresponding magnitudes having a numeric ratio close to the measured $H_{DMI}$ ratio from $Pt_1$ and $Pt_2$ wedged samples in series S1. Importantly, when simultaneously taking $\delta_1$ & $\delta_2$ into calculation, the resultant $\boldsymbol{D}$'s relative magnitude is almost identical to $\boldsymbol{D}(\delta_1) + \boldsymbol{D}(\delta_2)$ (Fig. S1(b)), a clear demonstration of additivity corresponding to the counter-wedge scenario where both displacements are effectively exploited. As shown in Fig S1(b), all combinations reminiscent of the complete four scenarios discussed in Figure 1d can be qualitatively assembled. For the SAF samples, this add-up rule also correctly predicts the relative strengths and overall ***D*** directions, as experimentally discovered in Figure 2.

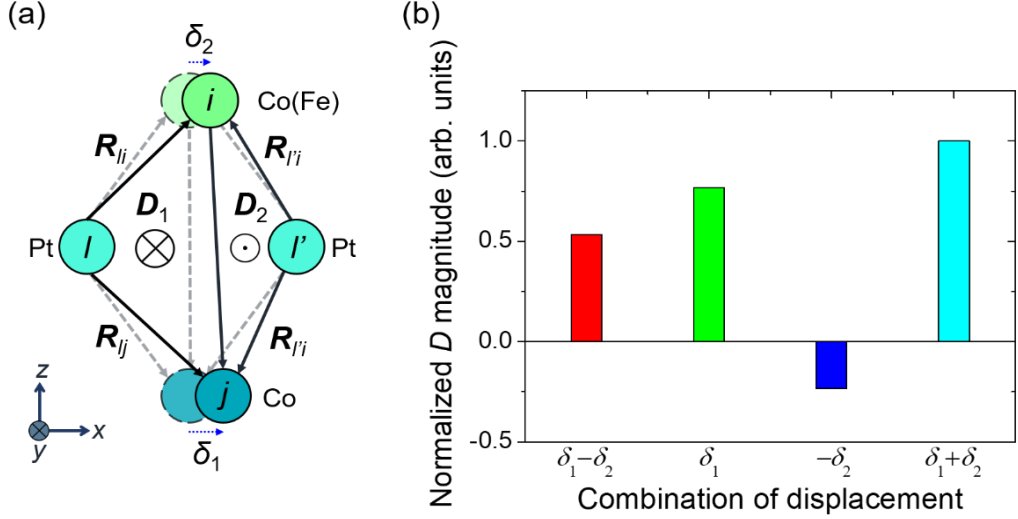

Fig. S1 (a) A toy model with two connected three-site triangles. The length between the nearest atoms is set to 2.78 Å (lattice constant of Pt(111)). The displacement $\delta_1$ and $\delta_2$ are set as 8% and 2.4% of the interatomic distance, respectively. (b) Normalized ***D*** magnitude under different displacement combinations, reminiscent to Figure 1d in the main text.

**Note II. Thickness gradient and out-of-plane anisotropy field across a distance**

We often emphasize the significance of columnar structures/template produced by OAD rather than the thickness gradient in the Pt layers due to the diminishingly small thickness difference in individual devices. For example, a previous work[56] reported $8.3 \times 10^{-4}$ nm thickness difference in a 5-µm wide device channel. Similarly, we calculated the Pt growth rate gradient produced by our own deposition system to be around $6.5 \times 10^{-7}$ (nm/s)/µm by measuring sputter rate at different substrate positions along the oblique deposition direction, or 0.16 nm thickness difference across a sizable substrate of, say, 5 cm in width. These are negligibly low values, and thus a thickness gradient itself is unlikely to generate interlayer DMI. Though, the thickness difference likely becomes non-negligible when upsized to wafer-scale as noted in the main text. We further investigate out-of-plane anisotropy fields ($H_k$) of two separate type-T devices (buffer Pt wedged) across a ~ 2.5 cm distance. As shown in Fig. S2(a) and (b), $H_k$ = 4277 ± 142 Oe and 4285 ± 173 Oe for the leftmost and rightmost device, respectively. The minor discrepancy of the measured anisotropies indicates limited impact of OAD on the magnetic properties.

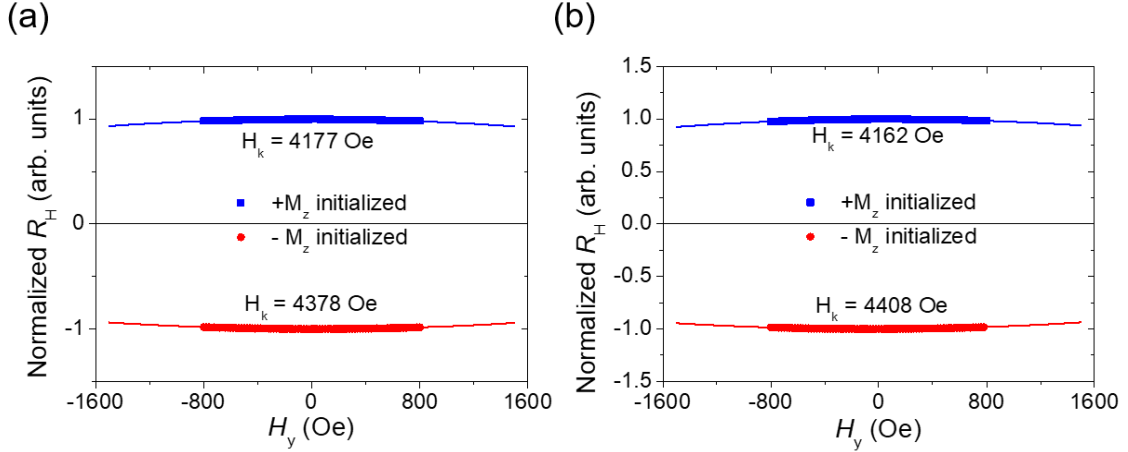

Fig. S2. Comparison of PMA anisotropy fields of two different devices separated by a ~ 2.5 cm distance. (a) and (b) captures $R_H$ response under y-directional field scan, in which $H_k$ is extracted by quadratic fitting. Results show good agreement between spatially separated devices.

**Note III. OAD incident angle dependence of $H_{DMI}$**

In the main text, the counter-wedge method to increase the overall $H_{DMI}$ is introduced. Another important point remains uninvestigated, namely changing the sputter incident angle $\theta_{incident}$ of the wedge layers. Schematics in Fig. S3(a) show the configuration between Pt target, $\theta_{incident}$ and the substrate. Fig. S3(b) presents the measured $H_{DMI}$ of type-T samples. The structures are similar to S1 but with various Pt spacer thicknesses ranges from 1.8 nm to 3.3 nm, and $\theta_{incident}$ altered from 20° to 70°. A mostly monotonically increasing $H_{DMI}$ is observed with elevated $\theta_{incident}$, with $H_{DMI}$ reaching as high as 285 Oe under low Pt spacer thickness of 1.8 nm and high $\theta_{incident}$ = 70°, one of the higher values reported in i-DMI systems. Similar to the case in Supplementary Information I and qualitatively mentioned in a previous work[26], increase of $H_{DMI}$ is accompanied by atomic in-plane shift's increase. We suspect the intensity of symmetry breaking is increased through elevated $\theta_{\text{incident}}$, bringing about a similar scenario in our devices resulting in increased $H_{\text{DMI}}$, in good agreement with the prediction.

On the other hand, since DMI is considered an additional term in the RKKY interaction[35], a damped oscillatory $H_{\text{DMI}}$ as a function of the spacer thickness is expected[26]. Indeed, a group of results in Fig. S3(b) under $\theta_{\text{incident}} = 50°$ captures the damped oscillatory behavior as the Pt spacer thickness increases, as individually shown in Fig. S3(c). However, no other obvious oscillatory behavior is found in other incident angles, which is attributed to the rather thick Pt spacer thickness (significant oscillatory behavior previously reported to occur[26] when the spacer thickness is predominantly less than ~ 1.5 nm) and further decreasing the spacer thickness might lead to more deterministic oscillations. Samples with SAF structures similar to S2 are also tested, with various $\theta_{\text{incident}}$ as stated above. The spacer thickness is unchanged to preserve the SAF characteristic, and focus is placed on the $H_{\text{DMI}}$-$\theta_{\text{incident}}$ behavior. Fig. S3(c) shows SAF structure has a large enhancement in $H_{\text{DMI}}$ of over four times by changing $\theta_{\text{incident}}$ from 20° to 70°, very similar to its type-T counterpart. Alongside the counter-wedge focused in the main text, changing the incident angle also serves well to increase $H_{\text{DMI}}$, which can be taken into consideration when designing future spintronic devices utilizing i-DMI.

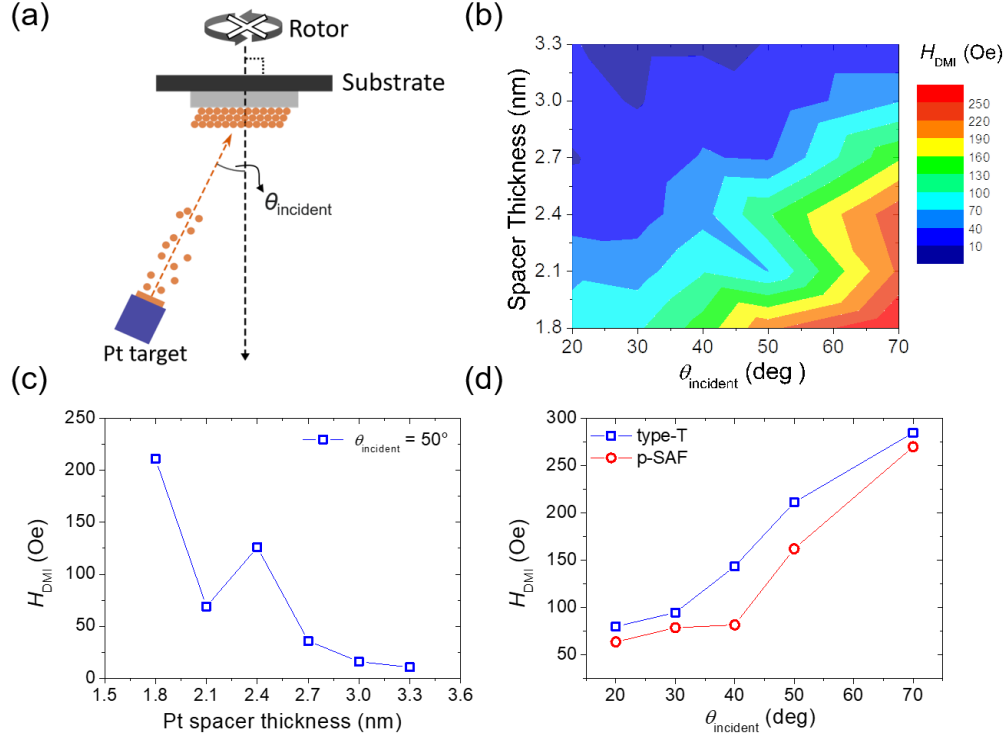

Fig. S3. Relations between the incident angle and $H_{DMI}$. (a) Schematics of the incident angle $\theta_{incident}$ with regard to the sample substrate. $\theta_{incident}$ ranged from 0° to 70° (b) Contour plot of type-T sample's $H_{DMI}$ as a function of Pt spacer thickness and $\theta_{incident}$. (c) Oscillatory interlayer DMI behavior extracted from the contour plot with $\theta_{incident} = 50°$. (d) $H_{DMI}$ vs. $\theta_{incident}$ for both type-T and SAF structures, the spacer Pt thickness of the type-T structure is 1.8 nm. Following the established protocol in the main text, $|H_{in}|$ during angle dependent loop-shift measurement is fixed at 200 Oe. All samples are grown with the counter-wedge technique (see Figure 1a and Figure 2a).

**Note IV. Damping-like SOT characterization, thermal stability and switching probability characterization for the SAF sample**

The effective fields produced by the current-induced damping-like SOT (DL-SOT) are characterized by the hysteresis loop shift method[38]. The two representative samples (type-T and SAF) presented here are identical to the ones used to measure field-free switching in Figure 3. As shown in Fig. S4(a), the extracted $H_z^{eff}/I_{DC}$ of the type-T sample indicates that the PMA magnetization experiences a negligible DL-SOT response of $H_z^{eff}/I_{DC} \leq 1.0$ Oe/mA, in line with the fact that the two

Pt layers with identical thicknesses cancel out the overall spin current. A much larger $H_z^{\text{eff}}/I_{\text{DC}} \sim 13.1$ Oe/mA is observed for the SAF sample under sufficiently high x-direction field $H_x$. Note that in the case of type-T devices, field-free switching can be achieved entirely relying on the i-DMI assist field due to the in-plane magnetization reversal of the CoFeB layer, without the Co layer experiencing a sizable SOT[41], of which the contribution cannot be detected by the current-induced hysteresis-loop shift method.

Thermal stability factor ($\Delta$) is then measured by field-free switching to test the robustness of the device. Under the thermally-activated switching scenario (due to the long applied current pulse width ($t_{\text{pulse}}$)), equation $I_c = I_{c0}\left[1-\frac{1}{\Delta}\ln(\frac{t_{\text{pulse}}}{\tau_0})\right]$ is employed[57], from which the zero thermal critical switching current, $I_{c0}$, and the thermal stability $\Delta$ can be extracted, with $\tau_0 \approx 1$ ns describing the intrinsic attempt rate. In practice, by varying $t_{\text{pulse}}$ from 200 μs to 2 ms and recording the corresponding switching current $I_c$, $I_{c0}$ and $\Delta$ are extracted by linearly fitting $I_c$ to $\ln(\frac{t_{\text{pulse}}}{\tau_0})$, as shown in Fig. S4(b). $\Delta = 24.9$ is insufficient for long-term data storage functionalities ($\Delta \geq 45$ required for a 10-year information retention time[58]), but still comparable to other contemporary reports on Hall bar devices[59, 60]. This may be improved by further stack engineering in the future.

Current pulses with opposite polarities are injected into the SAF sample ($\varphi_D = 90°$ device) with 50 ms pulse width and a current magnitude of $I_{\text{write}} = \pm 6$ mA to test the field-free operation's repeatability. Robust reversible switching behavior is successfully demonstrated, shown in Fig. S4(c), by treating a single switching's corresponding normalized $R_H \geq \pm 0.8$ as a successful event. Following an identical procedure, the switching percentage under different $I_{\text{write}}$ is plotted in Fig. S4(d), with a control sample showcasing failed field-free switching in a non-wedged device.

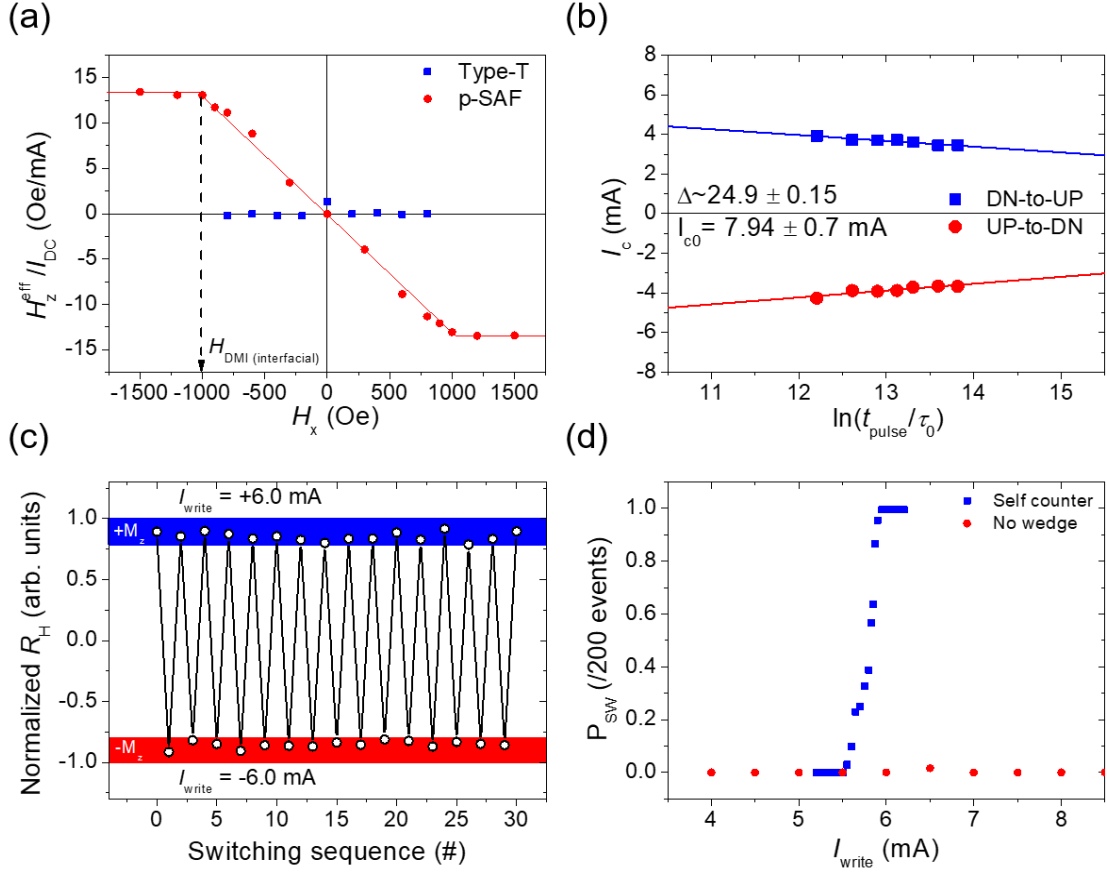

Fig. S4 Results of damping-like effective field characterizations and current-induced switching behaviors. (a) Strength of DL-SOT effective fields experienced by the PMA magnetization characterized by $H_z^{eff}/I_{DC}$. Saturation field (interfacial $H_{DMI}$) of the SAF sample labeled with black arrow. (b) Pulse width dependent measurement of critical switching current $I_c$. (c) Repeated switching events of the self-counter-wedged SAF device under $I_{write}$ with alternating polarity. The initial 30 sequences are shown from the overall 300 sequences. (d) Comparison of switching probability ($P_{sw}$) as a function of $I_{write}$ between the self-counter-wedged SAF sample, and a device prepared without wedge.

**Note V. $H_{DMI}$ characterizations and current-induced field-free switching in a type-T self-counter-wedged sample**

For a type-T sample, schematics of the "self-counter wedge" deposition scheme is illustrated in Fig. S5(a). Following the exhibition convention established in Figure 1 in the main text, blue/red stripes indicate the antiparallel deposition directions due to OAD and reversed OAD employed on the lower/upper half of the bottom Pt layer. This approach leads to a weakened but nevertheless still sizable $H_{\text{DMI}}$ of 51.8 Oe (Fig. S5(b)), as mentioned in the main text. The angle dependence of $\varphi_D \approx 280°$ is antiparallel compared to the counter-wedge scenario investigated in Figure 1 due to the opposite sequence of OAD and reversed OAD employed here (see the opposite color sequence of the stripes). Similarly, the $H_{\text{offset}} - \varphi_H$ data of the self-counter wedged SAF sample as shown in Fig. S4(b) shows $\varphi_D \approx 98°$ and $H_{\text{DMI}}$ of 51.8 Oe. Switching phenomena of said sample is also addressed in the main text.

The tested field-free switching loops of the type-T self-counter wedged sample are compiled in Fig. S5(c). It's obvious that failed switching occurs at devices with $\varphi_D = 90°$ and 270°, in agreement with the results reported in the main text, and a high switching percentage of ≈ 80% can be achieved in 0°/180° devices. However, the switching loops can show a "switch back" behavior, where the deterministic state of $R_H$ cannot be maintained under high current. This phenomenon has been observed in other works[24], and we attribute this phenomenon to the lowered $H_{\text{DMI}}$ causing a degraded symmetry breaking effect.

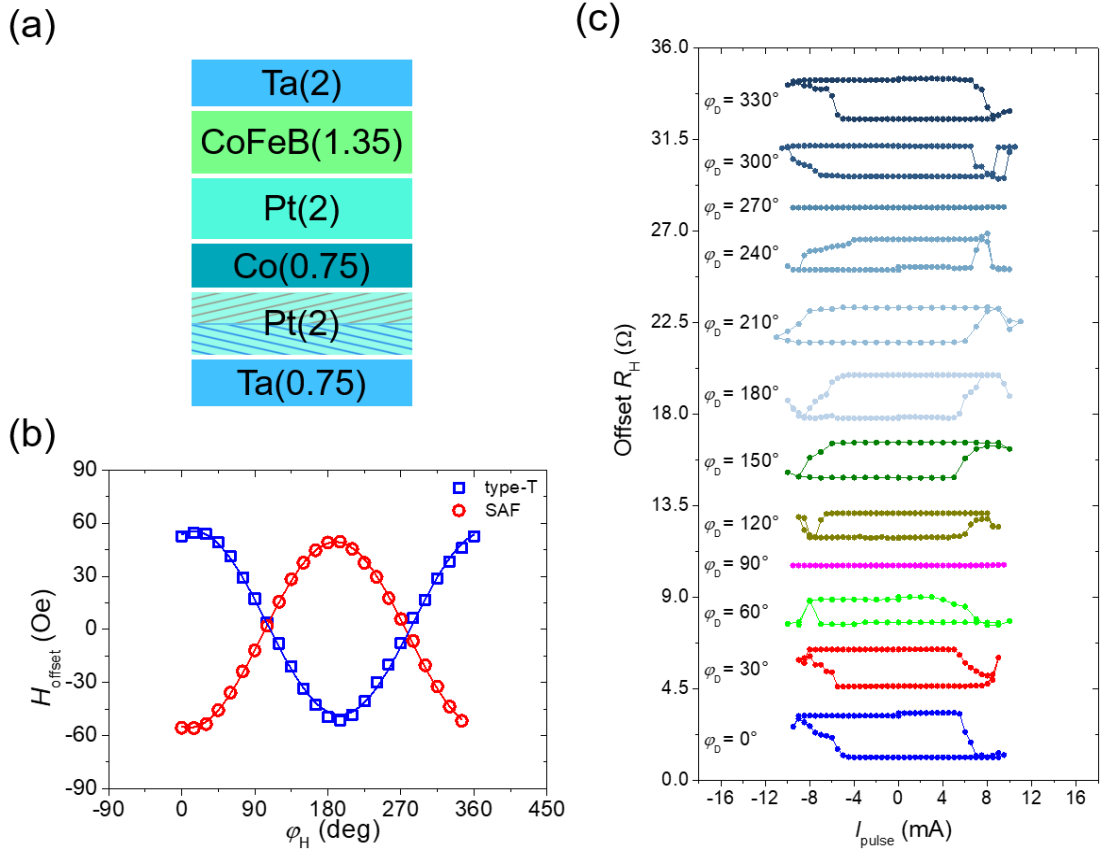

Fig S5. (a) Illustration of the structure and the self-counter-wedge condition in a device with type-T configuration. (b) Angle dependent loop shift of type-T and SAF devices, both grown with self-counter-wedge. (See Figure 4 in main text). (c) Switching curves of the type-T self-counter-wedged sample, with different $D$ vector directions ($\varphi_D$). While field-free switching remains achievable, there's obvious "switch-back" behaviors, likely due to the reduced $H_{DMI}$.

**Note VI. Individual layer's contribution in the self-counter wedge scenario, and weak $H_{DMI}$ from wedged Pt layers in the top FM layer**

The design of the self-counter wedged SAF layer stacks introduced in the main text is based on our model's predictions that owing to the difference in templation efficacy, interlayer DMI is in a self-counter wedged layer is dictated by the polarity of the lower half Pt sublayer, and that contributions from separate layers are multiplicative. To thoroughly addresses the self-counter wedge scenario, we further prepared samples where the overall self-counter wedge deposition (Figure 4a) is intentionally fragmented into three sections, namely $Pt_1$, $Pt_2$ and $Pt_3$. Their angle dependent loop-shift results, as shown in Fig. S6(a) indicate decisively that the three samples have identical ***D*** directions and align parallel to that described in Figure 2d of the main text. Notably, the combinatory nature of interlayer DMI is evident here, with the contributions from $Pt_1$, $Pt_2$ and $Pt_3$ add up to 62.5 Oe, fairly close to the reported $H_{DMI}$ = 52.5 Oe from the combined structure in Figure 4 and Supporting Information V, in line with our predictions. Wedged Pt layers in the top FM layer, on the other hand, contributes little to the overall interlayer DMI strength, with their corresponding $H_{DMI}$ negligibly small at around 5.6 Oe (Fig. S6(b)). This result echoes our observation that higher-level (later-grown) layers contribute much less to the interlayer DMI while compared to the bottom-level (earlier-grown) layers.

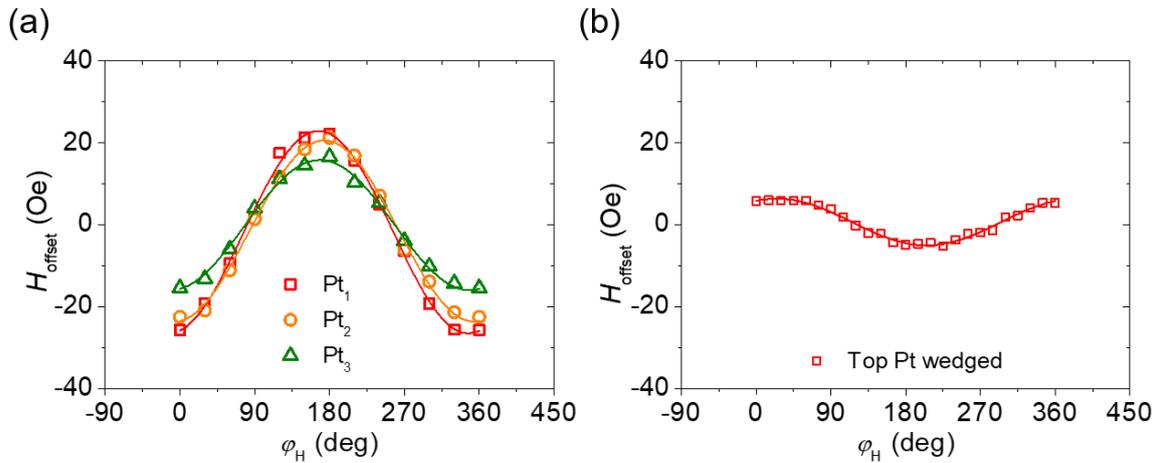

Fig. S6. Investigation of individual layer's contribution in the self-counter wedge scenario. (a) Contributions of self-counter wedge employed on $Pt_1$, $Pt_2$ and $Pt_3$ layers with $H_{DMI}$ magnitudes being 24.6 Oe, 22.0 Oe and 15.9 Oe, respectively. Note their constructive relationship and parallel alignment of ***D*** with regard to the case reported in Supporting Information V. (b) Angle scan for $H_{offset}$ when the top Pt layers are wedge deposited instead. Note the vastly weaker $H_{DMI}$ of 5.7 Oe.

# References

(1) Fukami, S.; Anekawa, T.; Zhang, C.; Ohno, H. A spin-orbit torque switching scheme with collinear magnetic easy axis and current configuration. *Nat. Nanotechnol.* **2016,** 11, (7), 621.

(2) Wolf, S. A.; Awschalom, D. D.; Buhrman, R. A.; Daughton, J. M.; von Molnár, S.; Roukes, M. L.; Chtchelkanova, A. Y.; Treger, D. M. Spintronics: A Spin-Based Electronics Vision for the Future. *Science* **2001,** 294, (5546), 1488-1495.

(3) Prenat, G.; Jabeur, K.; Vanhauwaert, P.; Pendina, G. D.; Oboril, F.; Bishnoi, R.; Ebrahimi, M.; Lamard, N.; Boulle, O.; Garello, K.; Langer, J.; Ocker, B.; Cyrille, M. C.; Gambardella, P.; Tahoori, M.; Gaudin, G. Ultra-Fast and High-Reliability SOT-MRAM: From Cache Replacement to Normally-Off Computing. *IEEE Trans. Multi-Scale Comput. Syst.* **2016,** 2, (1), 49-60.

(4) Krizakova, V.; Perumkunnil, M.; Couet, S.; Gambardella, P.; Garello, K. Spin-orbit torque switching of magnetic tunnel junctions for memory applications. *J. Magn. Magn. Mater.* **2022,** 562, 169692.

(5) Lee, D.; Go, D.; Park, H. J.; Jeong, W.; Ko, H. W.; Yun, D.; Jo, D.; Lee, S.; Go, G.; Oh, J. H.; Kim, K. J.; Park, B. G.; Min, B. C.; Koo, H. C.; Lee, H. W.; Lee, O.; Lee, K. J. Orbital torque in magnetic bilayers. *Nat. Comm.* **2021,** 12, (1), 6710.

(6) Hu, C.-Y.; Chiu, Y.-F.; Tsai, C.-C.; Huang, C.-C.; Chen, K.-H.; Peng, C.-W.; Lee, C.-M.; Song, M.-Y.; Huang, Y.-L.; Lin, S.-J.; Pai, C.-F. Toward 100% Spin–Orbit Torque Efficiency with High Spin–Orbital Hall Conductivity Pt–Cr Alloys. *ACS Appl. Electron. Mater.* **2022,** 4, (3), 1099-1108.

(7) Ryu, J.; Thompson, R.; Park, J. Y.; Kim, S.-J.; Choi, G.; Kang, J.; Jeong, H. B.; Kohda, M.; Yuk, J. M.; Nitta, J.; Lee, K.-J.; Park, B.-G. Efficient spin–orbit torque in magnetic trilayers using all three polarizations of a spin current. *Nat. Electron.* **2022,** 5, 217-223.

(8) Ikeda, S.; Miura, K.; Yamamoto, H.; Mizunuma, K.; Gan, H. D.; Endo, M.; Kanai, S.; Hayakawa, J.; Matsukura, F.; Ohno, H. A perpendicular-anisotropy CoFeB-MgO magnetic tunnel junction. *Nat. Mater.* **2010,** 9, (9), 721-4.

(9) van den Brink, A.; Vermijs, G.; Solignac, A.; Koo, J.; Kohlhepp, J. T.; Swagten, H. J. M.; Koopmans, B. Field-free magnetization reversal by spin-Hall effect and exchange bias. *Nat. Comm.* **2016,** 7, 10854.

(10) Nishimura, N.; Hirai, T.; Koganei, A.; Ikeda, T.; Okano, K.; Sekiguchi, Y.; Osada, Y. Magnetic tunnel junction device with perpendicular magnetization films for high-density magnetic random access memory. *J. Appl. Phys.* **2002,** 91, (8), 5246-5249.

(11) Miron, I. M.; Garello, K.; Gaudin, G.; Zermatten, P.-J.; Costache, M. V.; Auffret, S.; Bandiera, S.; Rodmacq, B.; Schuhl, A.; Gambardella, P. Perpendicular switching of a single ferromagnetic layer induced by in-plane current injection. *Nature* **2011,** 476, (7359), 189-193.

(12) Liu, L.; Pai, C.-F.; Li, Y.; Tseng, H. W.; Ralph, D. C.; Buhrman, R. A. Spin-Torque Switching with the Giant Spin Hall Effect of Tantalum. *Science* **2012,** 336, (6081), 555-558.

(13) P, V. M.; Ganesh, K. R.; Kumar, P. S. A. Spin Hall effect mediated current-induced deterministic switching in all-metallic perpendicularly magnetized Pt/Co/Pt trilayers. *Phys. Rev. B* **2017,** 96, (10), 104412.

(14) Yu, G.; Upadhyaya, P.; Fan, Y.; Alzate, J. G.; Jiang, W.; Wong, K. L.; Takei, S.; Bender, S. A.; Chang, L. T.; Jiang, Y.; Lang, M.; Tang, J.; Wang, Y.; Tserkovnyak, Y.; Amiri, P. K.; Wang, K. L. Switching of perpendicular magnetization by spin-orbit torques in the absence of external magnetic fields. *Nat. Nanotechnol.* **2014,** 9, (7), 548-54.

(15) Chen, R.; Cui, Q.; Liao, L.; Zhu, Y.; Zhang, R.; Bai, H.; Zhou, Y.; Xing, G.; Pan, F.; Yang, H.; Song, C. Reducing Dzyaloshinskii-Moriya interaction and field-free spin-orbit torque switching in synthetic antiferromagnets. *Nat. Comm.* **2021,** 12, (1).

(16) Ryu, J.; Avci, C. O.; Song, M.; Huang, M.; Thompson, R.; Yang, J.; Ko, S.; Karube, S.; Tezuka, N.; Kohda, M.; Kim, K. J.; Beach, G. S. D.; Nitta, J. Deterministic Current-Induced Perpendicular Switching in Epitaxial Co/Pt Layers without an External Field. *Adv. Funct. Mater.* **2023,** 33, (35).

(17) Oh, Y. W.; Chris Baek, S. H.; Kim, Y. M.; Lee, H. Y.; Lee, K. D.; Yang, C. G.; Park, E. S.; Lee, K. S.; Kim, K. W.; Go, G.; Jeong, J. R.; Min, B. C.; Lee, H. W.; Lee, K. J.; Park, B. G. Field-free switching of perpendicular magnetization through spin-orbit torque in antiferromagnet/ferromagnet/oxide structures. *Nat. Nanotechnol.* **2016,** 11, (10), 878-884.

(18) Lau, Y. C.; Betto, D.; Rode, K.; Coey, J. M.; Stamenov, P. Spin-orbit torque switching without an external field using interlayer exchange coupling. *Nat. Nanotechnol.* **2016,** 11, (9), 758-62.

(19) Fukami, S.; Zhang, C.; DuttaGupta, S.; Kurenkov, A.; Ohno, H. Magnetization switching by spin-orbit torque in an antiferromagnet-ferromagnet bilayer system. *Nat Mater* **2016,** 15, (5), 535-41.

(20) Kong, W. J.; Wan, C. H.; Wang, X.; Tao, B. S.; Huang, L.; Fang, C.; Guo, C. Y.; Guang, Y.; Irfan, M.; Han, X. F. Spin-orbit torque switching in a T-type magnetic configuration with current orthogonal to easy axes. *Nat. Comm.* **2019,** 10, (1), 233.

(21) Liu, L.; Zhou, C.; Shu, X.; Li, C.; Zhao, T.; Lin, W.; Deng, J.; Xie, Q.; Chen, S.; Zhou, J.; Guo, R.; Wang, H.; Yu, J.; Shi, S.; Yang, P.; Pennycook, S.; Manchon, A.; Chen, J. Symmetry-dependent field-free switching of perpendicular magnetization. *Nat. Nanotechnol.* **2021,** 16, (3), 277-282.

(22) Baek, S. C.; Amin, V. P.; Oh, Y. W.; Go, G.; Lee, S. J.; Lee, G. H.; Kim, K. J.; Stiles, M. D.; Park, B. G.; Lee, K. J. Spin currents and spin-orbit torques in ferromagnetic trilayers. *Nat Mater* **2018,** 17, (6), 509-513.

(23) Kateel, V.; Krizakova, V.; Rao, S.; Cai, K.; Gupta, M.; Monteiro, M. G.; Yasin, F.; Sorée, B.; De Boeck, J.; Couet, S.; Gambardella, P.; Kar, G. S.; Garello, K. Field-Free Spin–Orbit Torque Driven Switching of Perpendicular Magnetic Tunnel Junction through Bending Current. *Nano Lett.* **2023,** 23, (12), 5482-5489.

(24) Sun, C.; Jiao, Y.; Zuo, C.; Hu, X.; Tao, Y.; Jin, F.; Mo, W.; Hui, Y.; Song, J.; Dong, K. Field-free switching of perpendicular magnetization through spin-orbit torque in FePt/[TiN/NiFe](5) multilayers. *Nanoscale* **2021,** 13, (43), 18293-18299.

(25) Vedmedenko, E. Y.; Riego, P.; Arregi, J. A.; Berger, A. Interlayer Dzyaloshinskii-Moriya

Interactions. *Phys. Rev. Lett.* **2019,** 122, (25), 257202.

(26) Liang, S.; Chen, R.; Cui, Q.; Zhou, Y.; Pan, F.; Yang, H.; Song, C. Ruderman-Kittel-Kasuya-Yosida-Type Interlayer Dzyaloshinskii-Moriya Interaction in Synthetic Magnets. *Nano Lett.* **2023,** 23, (18), 8690-8696.

(27) Fernandez-Pacheco, A.; Vedmedenko, E.; Ummelen, F.; Mansell, R.; Petit, D.; Cowburn, R. P. Symmetry-breaking interlayer Dzyaloshinskii-Moriya interactions in synthetic antiferromagnets. *Nat. Mater.* **2019,** 18, (7), 679-684.

(28) Han, D. S.; Lee, K.; Hanke, J. P.; Mokrousov, Y.; Kim, K. W.; Yoo, W.; van Hees, Y. L. W.; Kim, T. W.; Lavrijsen, R.; You, C. Y.; Swagten, H. J. M.; Jung, M. H.; Klaui, M. Long-range chiral exchange interaction in synthetic antiferromagnets. *Nat. Mater.* **2019,** 18, (7), 703-708.

(29) Wang, K.; Qian, L.; Ying, S.-C.; Xiao, G. Spin-orbit torque switching of chiral magnetization across a synthetic antiferromagnet. *Commun. Phys.* **2021,** 4, (1), 10.

(30) He, W.; Wan, C.; Zheng, C.; Wang, Y.; Wang, X.; Ma, T.; Wang, Y.; Guo, C.; Luo, X.; Stebliy, M. E.; Yu, G.; Liu, Y.; Ognev, A. V.; Samardak, A. S.; Han, X. Field-Free Spin-Orbit Torque Switching Enabled by the Interlayer Dzyaloshinskii-Moriya Interaction. *Nano Lett.* **2022,** 22, (17), 6857-6865.

(31) Guo, Y.; Zhang, J.; Cui, Q.; Liu, R.; Ga, Y.; Zhan, X.; Lyu, H.; Hu, C.; Li, J.; Zhou, J.; Wei, H.; Zhu, T.; Yang, H.; Wang, S. Effect of interlayer Dzyaloshinskii-Moriya interaction on spin structure in synthetic antiferromagnetic multilayers. *Phys. Rev. B* **2022,** 105, (18), 184405.

(32) Avci, C. O.; Lambert, C. H.; Sala, G.; Gambardella, P. Chiral Coupling between Magnetic Layers with Orthogonal Magnetization. *Phys. Rev. Lett.* **2021,** 127, (16), 167202.

(33) Dzyaloshinsky, I. A thermodynamic theory of 'weak' ferromagnetism of antiferromagnetics. *J. Phys. Chem. Sol.* **1957,** 4, 241-255.

(34) Moriya, T. Anisotropic Superexchange Interaction and Weak Ferromagnetism. *Phys. Rev.* **1960,** 120, (1), 91-98.

(35) Fert, A.; Levy, P. M. Role of Anisotropic Exchange Interactions in Determining the Properties of Spin-Glasses. *Phys. Rev. Lett.* **1980,** 44, (23), 1538-1541.

(36) Bogdanov, A. N.; Rossler, U. K. Chiral symmetry breaking in magnetic thin films and multilayers. *Phys. Rev. Lett.* **2001,** 87, (3), 037203.

(37) Torrejon, J.; Kim, J.; Sinha, J.; Mitani, S.; Hayashi, M.; Yamanouchi, M.; Ohno, H. Interface control of the magnetic chirality in CoFeB/MgO heterostructures with heavy-metal underlayers. *Nat. Comm.* **2014,** 5, 4655.

(38) Pai, C.-F.; Mann, M.; Tan, A. J.; Beach, G. S. D. Determination of spin torque efficiencies in heterostructures with perpendicular magnetic anisotropy. *Phys. Rev. B* **2016,** 93, (14), 144409.

(39) Dao, T. P.; Muller, M.; Luo, Z.; Baumgartner, M.; Hrabec, A.; Heyderman, L. J.; Gambardella, P. Chiral Domain Wall Injector Driven by Spin-Orbit Torques. *Nano Lett.* **2019,** 19, (9), 5930-5937.

(40) Kammerbauer, F.; Choi, W. Y.; Freimuth, F.; Lee, K.; Fromter, R.; Han, D. S.; Lavrijsen, R.; Swagten, H. J. M.; Mokrousov, Y.; Klaui, M. Controlling the Interlayer Dzyaloshinskii-Moriya Interaction by Electrical Currents. *Nano Lett.* **2023,** 23, (15), 7070-7075.

(41) Huang, Y.-H.; Huang, C.-C.; Liao, W.-B.; Chen, T.-Y.; Pai, C.-F. Growth-Dependent Interlayer Chiral Exchange and Field-Free Switching. *Phys. Rev. Appl.* **2022,** 18, (3), 034046.
(42) Li, Y.-C.; Huang, Y.-H.; Huang, C.-C.; Liu, Y.-T.; Pai, C.-F. Field-Free Switching in Symmetry-Breaking Multilayers: The Critical Role of Interlayer Chiral Exchange. *Phys. Rev. Appl.* **2023,** 20, (2), 024036.
(43) Harris, K. D.; van Popta, A. C.; Sit, J. C.; Broer, D. J.; Brett, M. J. A Birefringent and Transparent Electrical Conductor. *Adv. Funct. Mater.* **2008,** 18, (15), 2147-2153.
(44) Taschuk, M. T.; Sorge, J. B.; Steele, J. J.; Brett, M. J. Ion-Beam Assisted Glancing Angle Deposition for Relative Humidity Sensors. *IEEE Sens. J.* **2008,** 8, (9), 1521-1522.
(45) Barranco, A.; Borras, A.; Gonzalez-Elipe, A. R.; Palmero, A. Perspectives on oblique angle deposition of thin films: From fundamentals to devices. *Prog. Mater. Sci.* **2016,** 76, 59-153.
(46) Arregi, J. A.; Riego, P.; Berger, A.; Vedmedenko, E. Y. Large interlayer Dzyaloshinskii-Moriya interactions across Ag-layers. *Nat. Comm.* **2023,** 14, (1), 6927.
(47) Parkin, S. S.; Mauri, D. Spin engineering: Direct determination of the Ruderman-Kittel-Kasuya-Yosida far-field range function in ruthenium. *Phys. Rev. B* **1991,** 44, (13), 7131-7134.
(48) Cuchet, L.; Rodmacq, B.; Auffret, S.; Sousa, R. C.; Prejbeanu, I. L.; Dieny, B. Perpendicular magnetic tunnel junctions with a synthetic storage or reference layer: A new route towards Pt- and Pd-free junctions. *Sci. Rep.* **2016,** 6, 21246.
(49) Wang, Z.; Li, P.; Fattouhi, M.; Yao, Y.; Van Hees, Y. L. W.; Schippers, C. F.; Zhang, X.; Lavrijsen, R.; Garcia-Sanchez, F.; Martinez, E.; Fert, A.; Zhao, W.; Koopmans, B. Field-free spin-orbit torque switching of synthetic antiferromagnet through interlayer Dzyaloshinskii-Moriya interactions. *Cell Rep.* **2023,** 4, (4), 101334.
(50) Hajihoseini, H.; Kateb, M.; Ingvarsson, S.; Gudmundsson, J. T. Oblique angle deposition of nickel thin films by high-power impulse magnetron sputtering. *Beilstein J. Nanotechnol* **2019,** 10, 1914-1921.
(51) Chuang, T. C.; Pai, C. F.; Huang, S. Y. Cr-induced Perpendicular Magnetic Anisotropy and Field-Free Spin-Orbit-Torque Switching. *Phys. Rev. Appl.* **2019,** 11, (6), 016005.
(52) Hu, C.-Y.; Chen, W.-D.; Liu, Y.-T.; Huang, C.-C.; Pai, C.-F. The Central Role of Tilted Anisotropy for Field-Free Spin-Orbit Torque Switching of Perpendicular Magnetization. *arXiv e-prints* **2023**, arXiv:2306.06357.
(53) Gao, F. S.; Liu, S. Q.; Zhang, R.; Xia, J. H.; He, W. Q.; Li, X. H.; Luo, X. M.; Wan, C. H.; Yu, G. Q.; Su, G.; Han, X. F. Experimental evidence of the oscillation behavior of the interlayer DMI effect. *Appl. Phys. Lett.* **2023,** 123, (19), 192401.
(54) Liu, Y.-T.; Huang, C.-C.; Chen, K.-H.; Huang, Y.-H.; Tsai, C.-C.; Chang, T.-Y.; Pai, C.-F. Anatomy of Type-x Spin-Orbit-Torque Switching. *Phys. Rev. Appl.* **2021,** 16, (2), 024021.
(55) Zhou, X. W.; Wadley, H. N. G. Atomistic simulation of the vapor deposition of Ni/Cu/Ni multilayers: Incident adatom angle effects. *J. Appl. Phys.* **2000,** 87, (1), 553-563.
(56) Masuda, H.; Seki, T.; Yamane, Y.; Modak, R.; Uchida, K.-i.; Ieda, J. i.; Lau, Y.-C.; Fukami, S.;

Takanashi, K. Large Antisymmetric Interlayer Exchange Coupling Enabling Perpendicular Magnetization Switching by an In-Plane Magnetic Field. *Phys. Rev. Appl.* **2022,** 17, (5), 054036.

(57) Koch, R. H.; Katine, J. A.; Sun, J. Z. Time-resolved reversal of spin-transfer switching in a nanomagnet. *Phys. Rev. Lett.* **2004,** 92, (8), 088302.

(58) Dieny, B.; Chshiev, M. Perpendicular magnetic anisotropy at transition metal/oxide interfaces and applications. *Rev. Mod. Phys.* **2017,** 89, (2).

(59) Kim, H. J.; Moon, K. W.; Tran, B. X.; Yoon, S.; Kim, C.; Yang, S.; Ha, J. H.; An, K.; Ju, T. S.; Hong, J. I.; Hwang, C. Field-Free Switching of Magnetization by Tilting the Perpendicular Magnetic Anisotropy of Gd/Co Multilayers. *Adv. Funct. Mater.* **2022,** 32, 2112561.

(60) Saito, Y.; Ikeda, S.; Endoh, T. Enhancement of current to spin-current conversion and spin torque efficiencies in a synthetic antiferromagnetic layer based on a Pt/Ir/Pt spacer layer. *Phys. Rev. B* **2022,** 105, (5).